%% file: 0_main.tex
\documentclass{article}
\usepackage{amsmath,graphicx}
\usepackage{graphicx}
\usepackage[group-separator={,}]{siunitx}

\usepackage{color}
\usepackage[table]{xcolor} 
\usepackage{subcaption}
\usepackage{tabularx}
\usepackage{booktabs}
\usepackage{float}
\usepackage{multirow, array}
\usepackage{comment}
\usepackage{hyperref}

\usepackage[preprint]{spconf}

\title{
    UniToPatho, a labeled histopathological dataset for colorectal polyps classification and adenoma dysplasia grading
}

\name{
    \begin{tabular}{c}
    Carlo Alberto Barbano$^\star$\qquad Daniele Perlo$^\star$\qquad 
    Enzo Tartaglione$^\star$\\ Attilio Fiandrotti$^\star$\qquad 
    Luca Bertero$^\dagger$\qquad Paola Cassoni$^\dagger$\qquad
    Marco Grangetto$^\star$\sthanks{This project has received funding from the European Union’s Horizon 2020 research and innovation programme under grant agreement No 825111, DeepHealth Project.}
    \end{tabular}
}

\address{
  $^{\star}$ Computer Science Department, University of Turin, 10149 Torino, Italy \\
  $^{\dagger}$ Medical Sciences Department, University of Turin, 10126, Torino, Italy
}

\begin{document}

\onecolumn
\noindent © 20XX IEEE. Personal use of this material is permitted. Permission from IEEE must be obtained for all other uses, in any current or future media, including reprinting/republishing this material for advertising or promotional purposes, creating new collective works, for resale or redistribution to servers or lists, or reuse of any copyrighted component of this work in other works.
\twocolumn
\clearpage

\maketitle
\ninept

\begin{abstract}
	Histopathological characterization of colorectal polyps allows to tailor patients' management and follow up with the ultimate aim of avoiding or promptly detecting an invasive carcinoma.
	Colorectal polyps characterization relies on the histological analysis of tissue samples to determine the polyps malignancy and dysplasia grade.
	Deep neural networks achieve outstanding accuracy in medical patterns recognition, however they require large sets of annotated training images.
	We introduce \textit{UniToPatho}, an annotated dataset of \num{9536} hematoxylin and eosin (H\&E) stained patches extracted from 292 whole-slide images,
	meant for training deep neural networks for colorectal polyps classification and adenomas grading.
	We present our dataset and provide insights on how to tackle the problem of automatic colorectal polyps characterization.
\end{abstract}


\input{1_introduction}

\input{3_dataset}
\input{4_analysis}
\input{5_ensemble}

\input{6_experiments}

\input{7_conclusion}

\bibliographystyle{IEEEbib}
\bibliography{col_bibliography}

\end{document}

%% file: 1_introduction.tex
\section{Introduction}
\label{sec:introduction}

The demand for gastrointestinal histopathology is on the rise~\cite{gonzalez2020updates}, fostered by the widespreading of cancer screening programs.
Gastrointestinal histopathologists inspect tissue samples collected during colonoscopies,
looking for hints that can predict the insurgence of invasive carcinoma~\cite{bevan2018colorectal}.
Colorectal polyps are pre-malignant lesions found in the intestinal mucosa that pathologysts analyze to \emph{i)} ascertain the polyp type (hyperplastic, adenoma) and \emph{ii)} assess the dysplasia grade in case of adenomas.
Examination of colorectal polyps represents a large share of histopathologists workload, thus methods for automating these tasks are highly sought.
Despite such clinical relevance, the concordance rate even among expert pathologists, in the diagnostic assessment of colorectal polyps, is far from optimal~\cite{denis2009diagnostic,mollasharifi2020interobserver}. Although the distinction between non-adenomatous and adenomatous tissue is usually reliable, the inter-observer agreement between different histological types and dysplasia grades is sub-optimal. For instance, the concordance in assessing a tubulo-villous polyp or low grade dysplasia ranged around 70\% \cite{denis2009diagnostic}.\

\noindent Deep learning-based methods have shown promising results towards automating the pathologists' work~\cite{Janowczyk16}. 
Korbar~\emph{et~al.}~\cite{korbar2017deep} present a patch-based framework, developed using a ResNet architecture~\cite{he2016deep}, to classify different types of colorectal polyps from whole-slide images. Their work provides empirical suggestion that residual architectures are better suited at this task.
Wei~\emph{et~al.}~\cite{wei2020evaluation} propose an analysis model for annotated tissue samples and perform a study on the generalization of neural models with external medical institutions.
Their work describes a hierarchical evaluation mechanism to extend the classification of tissue fragments to the entire slide.
Song~\emph{et~al.}~\cite{song2020automatic} propose a patch-based fully-convolutional approach for the classification and grading of adenomas, with a strong focus on model interpretability.
They also highlight how different patch sizes should be used for adenomas classification and grading. \\
However, the scarcity of datasets large enough and suitably labeled represents a major a hurdle for training deep-learning based algorithms to predict polyp type and adenoma dysplasia grade. 

This work provides the following contributions towards automatic colorectal polyps characterization, in the framework of the \textit{DeepHealth}~\cite{DeepHealth} project. 
\\
First, we make available \textit{UniToPatho}\footnote{The dataset is available at \url{https://ieee-dataport.org/open-access/unitopatho}}, a high-resolution annotated dataset of Hematoxylin and Eosin (H\&E)-stained colorectal images~\cite{unitopatho}.
UniToPatho enables training deep neural networks to classify  different colorectal polyps types and adenomas grading.
We make available our annotated dataset as a collection of high-resolution patches extracted at different scales.
\\
Second, we show that the direct application of a deep neural network fails to classify both the tissue type and adenoma dysplasia grade.\\
Lastly, we propose
a multi-resolution deep learning approach solving the previous issues, that achieves significant accuracy in the characterization of colorectal polyps.

%% file: 3_dataset.tex
\section{The UniToPatho dataset}
\label{sec:Dataset}

\addtolength{\tabcolsep}{-1pt}  
\begin{table}
    \centering
    \footnotesize
    \vspace{5pt}
    \begin{tabular}{l c c c c c c c}
        \toprule
         & \ \  HP\ \ \   & NORM & \multicolumn{2}{c}{TA} & \multicolumn{2}{c}{TVA} & \textbf{Total}\\
         & & & HG & LG & HG & LG & \\
        \midrule
        Slides & 41 & 21 & 26 & 146 & 20 & 38 & {\bfseries 292}\\
        \midrule
        $\sigma$ = 7000 & 59  & 74  & 98  & 411  & 93  & 132  & \textbf{867} \\
        $\sigma$ = 800 & 545 & 950 & 454 & 3618 & 916 & 2186 & \textbf{8699} \\
        Total & 604 & 1024 & 552 & 4029 & 1009 & 2318 & \textbf{9536} \\
        \bottomrule
    \end{tabular}
    \caption{\textit{UniToPatho} class distribution for whole image slides (top) and the two patch scales made available (bottom).}
    \label{tab:dataset-details}
\end{table}

\addtolength{\tabcolsep}{1pt}  
\begin{figure}
    \centering
    \begin{subfigure}{0.25\columnwidth}
      \includegraphics[width=\columnwidth]{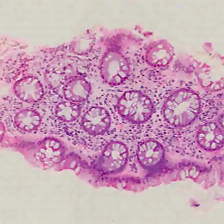}
      \caption{NORM}
    \end{subfigure}\hfil 
    \begin{subfigure}{0.25\columnwidth}
      \includegraphics[width=\columnwidth]{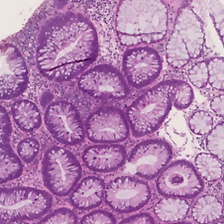}
      \caption{TA.LG}
    \end{subfigure}\hfil 
    \begin{subfigure}{0.25\columnwidth}
      \includegraphics[width=\columnwidth]{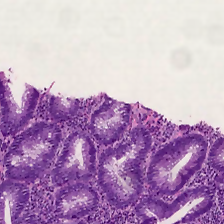}
      \caption{TA.HG}
    \end{subfigure}
    
    \medskip
    
    \begin{subfigure}{0.25\columnwidth}
      \includegraphics[width=\columnwidth]{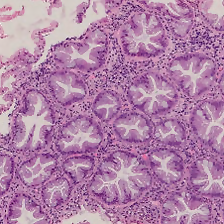}
      \caption{HP}
    \end{subfigure}\hfil 
    \begin{subfigure}{0.25\columnwidth}
      \includegraphics[width=\columnwidth]{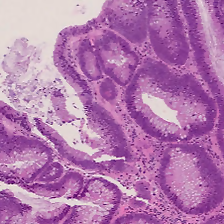}
      \caption{TVA.LG}
    \end{subfigure}\hfil
    \begin{subfigure}{0.25\columnwidth}
      \includegraphics[width=\columnwidth]{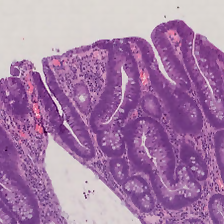}
      \caption{TVA.HG}
    \end{subfigure}
    \caption{Example of 800$\times$800\textmu m patches for the six UniToPatho colorectal polyps classes.}
    
    \label{fig:dataset-sample}
\end{figure}

\textit{UniToPatho} is a dataset of annotated high-resolution H\&E-stained images, comprising different histological samples of colorectal polyps, collected from patients undergoing cancer screening. The dataset is a collection of the most relevant patch images extracted from \num{292} \textit{whole-slide images} (simply \textit{slides} in the following), in accordance with UniTo pathologists' evaluation.
The slides are acquired through a Hamamatsu Nanozoomer S210 scanner at \num{20}$\times$ magnification (\SI{0.4415}{\micro\meter}/px), as exemplified in Fig.~\ref{fig:dataset-sample}.
Each slide belongs to a different patient and is annotated by expert UniTo pathologists,
according to six classes as follows:
\\

\noindent \textbf{NORM} - Normal tissue \\
\noindent \textbf{HP} - Hyperplastic Polyp \\
\noindent \textbf{TA.HG} - Tubular Adenoma, High-Grade dysplasia \\
\noindent \textbf{TA.LG} - Tubular Adenoma, Low-Grade dysplasia \\
\noindent \textbf{TVA.HG} - Tubulo-Villous Adenoma, High-Grade dysplasia \\
\noindent \textbf{TVA.LG} - Tubulo-Villous Adenoma, Low-Grade dysplasia \\

\noindent Hyperplastic polyps usually exhibit no malignant potential~\cite{Tseung2005RobbinsAC}, while adenomas are more likely to progress into invasive carcinomas.
Tubular and tubulo-villous are common colorectal adenomas, with villous adenomas generally presenting higher malignant potential given the larger surface~\cite{Tseung2005RobbinsAC}.
Adenomas are associated with a grade of dysplasia, which measures the abnormality in cellular growth and differentiation~\cite{NCI}. Higher grade dysplasia indicates higher malignant potential. 

We split the slides into a train set and a test set with a \num{70}\% to \num{30}\% ratio, resulting in \num{204} slides used for training and 88 for testing. 
Therefore, each slide is represented either in the train set or in the test set, but not in both. 
Following the approach in~\cite{song2020automatic, kather2018}, we release square patches cropped from the \num{292} slides.
From each slide, we crop multiple non-overlapping square patches at different scales.
We denote the side of the underlying physical area with $\sigma$, measured in \textmu m, which we vary from 100 to 8000.
The number of patches obtained for each slide hence depends on multiple factors including $\sigma$, the slide size and the polyp type. As common in similar datasets, the different classes are highly unbalanced in the dataset.
We make publicly available a total of \num{9536} patches, \num{8669} of which extracted at ${\sigma=\num{800}}$ (\num{1812}$\times$\num{1812} pixels patches) and \num{867} at ${\sigma=\num{7000}}$ (\num{15855}$\times$\num{15855} pixels patches). These are the sizes that will be used in Sec.~\ref{method:ensemble}.
Tab.~\ref{tab:dataset-details} provides a summary of the class distribution for the whole slides and the released patches. 

%% file: 4_analysis.tex
\section{Preliminary Analysis}
\label{method:preliminary}

We perform our preliminary experiments on the UniToPatho dataset with a baseline strategy owing to state-of-the-art methods~\cite{korbar2017deep,wei2020evaluation}.
The method which will be described in the next section builds upon the lessons learned during these experiments.

First, we randomly color-jitter the patches described in the previous section to augment the dataset, as proposed in~\cite{wei2020evaluation, korbar2017deep}.
Next, as in~\cite{korbar2017deep, wei2020evaluation}, we train a deep convolutional neural network for image classification on the patches belonging to the train slides.
Namely, we train an ImageNet-pretrained ResNet-18~\cite{he2016deep} with SGD for 50 epochs, with an initial learning rate of 0.01 decayed by a factor of 0.1 every 20 epochs.
The output layer is reshaped to match the UniToPatho 6-class classification problem.
Each patch is downsampled to the standard 224$\times$224 pixels ImageNet size prior being fed into the ResNet.
We repeat the whole training process for different scales, choosing $\sigma$ in the [100, 8000] range.
For each $\sigma$, the trained network is eventually tested on the patches extracted at the same scale from the test slides.

\begin{table}
    \centering
    \begin{tabular}{r | c c c c c c}
        \toprule
        & \multicolumn{6}{c}{Patch scale $\sigma$ [\textmu m]}\\
        \midrule
        Type & 100 & 800 & 1500 & 4000 & 7000 & 8000 \\
        \midrule
        BA (6-class) & 0.40 & 0.45 & \textbf{0.46} & 0.41 & 0.37 & 0.38 \\
        \midrule
        NORM & 0.70 & 0.66 & 0.72 & 0.76 & 0.78 & 0.71\\
        HP &  0.81 & {\bfseries 0.92} & 0.85 & 0.70 & 0.60  & 0.69 \\
        TA (HG+LG) & 0.65 & 0.66  & 0.65 & 0.71 & {\bfseries 0.76} & 0.70 \\
        TVA (HG+LG) & 0.64 & 0.67 & 0.68 & 0.74 & {\bfseries 0.84} & 0.76 \\
        \bottomrule
    \end{tabular}
    \caption{Preliminary experiments: overall BA for all of the six classes (first row) and BA for each polyp type, plus normal tissue.}
    \label{tab:class-ba}
\end{table}

We analyze the the results of our preliminary experiments first in terms of polyp type classification accuracy, next in terms of adenoma grade prediction for adenoma samples only. 
In the following, the classification accuracy will be defined in terms of Balanced Accuracy (BA) to cope with the unbalanced samples in the dataset.
In Tab.~\ref{tab:class-ba} (first row) we show the BA achieved when attempting to discriminate all 6 classes with the baseline approach as function of the patch scale $\sigma$: it can be noted that even the best accuracy achieved at $\sigma$=1500 is quite low. 
Nonetheless, we conjecture that different polyp types can be better recognized at different scales: as a consequence, in Tab.~\ref{tab:class-ba} we also 
show the BA concerning the classification of each single polyp type 
HP, TA, TVA plus NORM.
The TA and TVA classes encompass the respective low grade (LG) and high grade (HG) subclasses.
Indeed, breaking down the accuracy on a per-class basis reveals that different types of polyps achieve top-classification accuracy at different scales.

Hyperplastic Polyps (HP) are best classified at a finer 800\textmu m scale: we hypothesize these types of benign polyps are best discriminated by looking at smaller-scale details such as gland edges~\cite{taherian2020tubular}.
Conversely, Tubular Adenomas (TA) and  Tubulo-Villous Adenomas (TVA) are best classified at a coarser 7000\textmu m scale: we hypothesize this type of polyps is best discriminated by looking 
at large-scale macro structures such as entire glands shapes~\cite{mehtat2020moleculary}. 

Coming to the problem of predicting the grade (LG or HG) of TA and TVA adenomas, we investigate whether that could be best predicted at some particular scale $\sigma$.
Our experiments proved however inconclusive, i.e. the adenoma grade classification accuracy does not appear being a function of the patch scale.
In fact, visual inspection of the 224$\times$224px downsampled patches reveals that the downsampling trashes discriminative details in the cells nuclei upon which pathologists are known to rely.

Concluding, our preliminary experiments show that \emph{i)} different polyp types are best classified at different scales (in our case ${\sigma=800}$ for HP and ${\sigma=7000}$ for TA and TVA) and \emph{ii)} adenoma grade prediction may be jeopardized if the details in cells nuclei are lost.
These findings are exploited to devise our proposed method detailed in the next section.

%% file: 5_ensemble.tex
\section{Proposed Method}
\label{method:ensemble}

\begin{figure}
    \centering
    \includegraphics[width=\columnwidth]{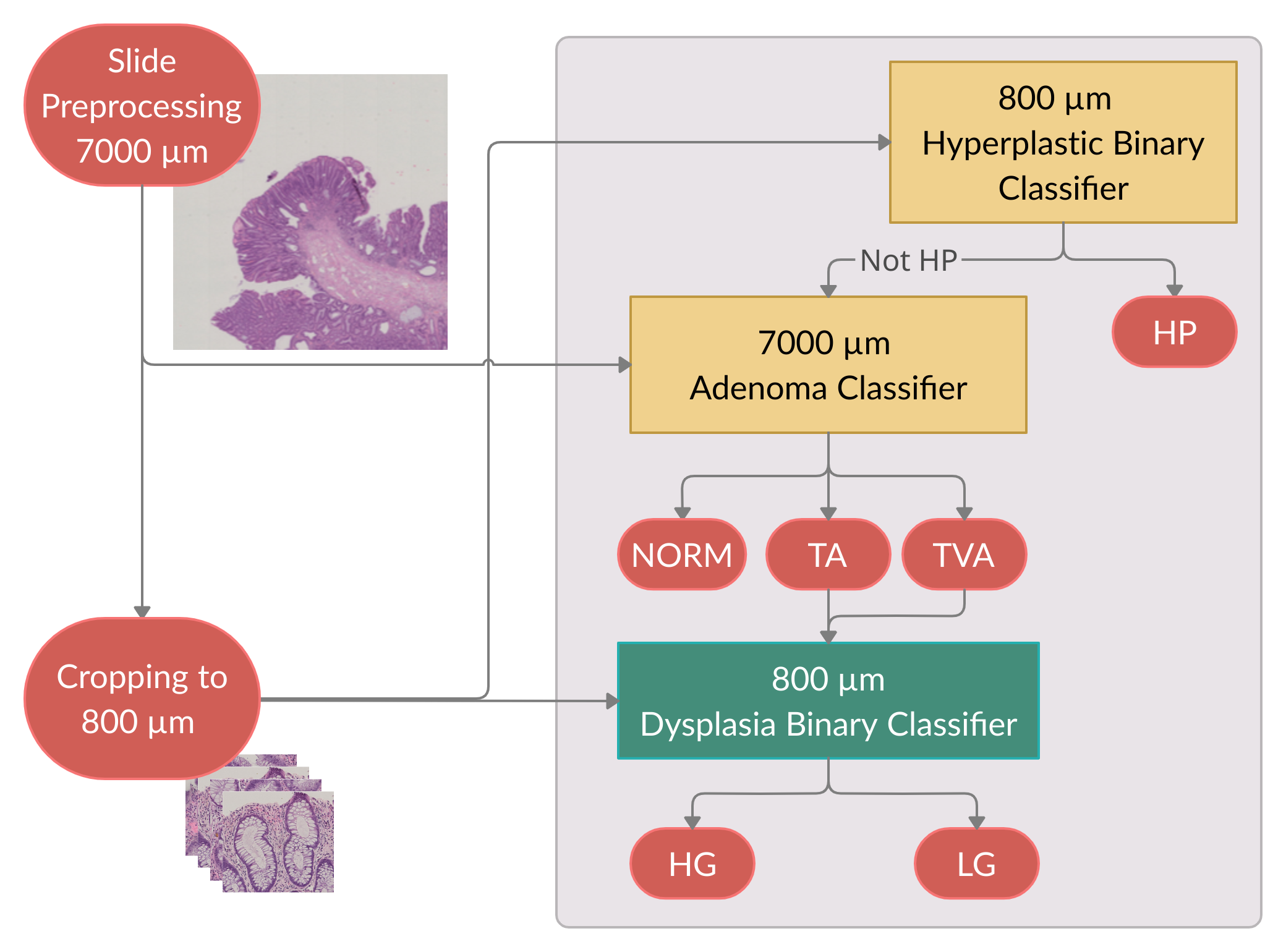}
    \caption{The proposed multi-resolution ensemble of cascaded classifiers. The Hyperplastic and Adenoma classifiers (yellow) are trained with inputs downsampled to 224$\times$224 pixels, while the Dysplasia classifier (green) uses full resolution images.}
    \label{fig:method_schema}
\end{figure}

This section details our proposed approach towards classification of UniToPatho images: a multi-resolution ensemble of cascaded classifiers. 
The method relies on three cascaded ResNet-18 classifiers, having the output layer specifically adapted for each classification task.
The classifiers are trained on patches extracted at either ${\sigma=800}$ or ${\sigma=7000}$, following the procedure described in the previous section.

The overall process of inference is depicted in Fig.~\ref{fig:method_schema}, with the input being a single 7000$\times$7000\textmu m patch which is used by the three ResNet-18 classifiers mentioned above.
For classifiers working at ${\sigma=800}$, we crop the input image into smaller 800$\times$800\textmu m sub-patches which are used to generate a prediction for the entire image.
Also, all of the patches are downsampled to 224$\times$224 pixels, unless stated otherwise.

\subsection{HP polyps detection}

First, HP polyps are discriminated from adenomas and normal images via a binary classifier that takes as input sub-patches of size 800$\times$800\textmu m extracted from the larger input image.
In fact, Tab.~\ref{tab:class-ba} shows that HP polyps are identified with top accuracy (0.92 BA)  at scale ${\sigma=800}$.
To infer the probability of predicting HP for entire image, we compute the average predicted probability of the sub-patches of being HP.
In the case the patch content is not classified as an HP polyp, the second classifier in the cascade is invoked.

\subsection{Adenoma detection}

Second, TA adenomas are discriminated from TVA adenomas and from normal images via a ternary classifier taking as input the entire 7000$\times$7000\textmu m patch.
In fact, Tab.~\ref{tab:class-ba} shows that TA and TVA adenomas are identified with top accuracy at this scale.
In the case the patch content is classified as a TA or TVA polyp, the third and last classifier is invoked to infer its grade, otherwise the tissue will be labeled as NORM.

\subsection{Dysplasia grading}

Finally, a binary classifier is used to determine the dysplasia grade for TA and TVA adenomas.
Sec.~\ref{method:preliminary} suggested that downsampling the patches to 224$\times$224px might be detrimental towards inferring the dysplasia grade due to the loss of important features such as cells nuclei, as also observed by~\cite{song2020automatic}.
Thus, only for this specific classification task, we skip downsampling to prevent the loss of fine grained details.
To account for the increased size of the feature space, we add an adaptive average pooling layer~\cite{he2015spatial, van2019evolutionary} before the fully connected layer of the ResNet-18 network.
We repeat the experiment of Sec.~\ref{method:preliminary} sweeping the patch scale $\sigma$ this time without downsampling and we find that adenoma grade classification peaks (0.81 BA) for ${\sigma=800}$.
As a consequence, we apply the grade classifier on sub-patches  extracted from the input image at the scale ${\sigma=800}$.
Finally, we infer the dysplasia grade for the entire input image according to a threshold $T_d$: if the ratio of high grade predictions on the sub-patches is higher than $T_d$, the image is labeled as HG, otherwise as LG. 
This choice is motivated by pathologists grading guidelines, where HG can be decided on the base of small portions of tissue; clearly, setting smaller values for $T_d$ can mimic this behaviour~\cite{Tseung2005RobbinsAC}.


\begin{figure*}
    \begin{minipage}[b]{.49\textwidth}
        \centering
        \begin{subfigure}{0.9\columnwidth}
            \centering
            \includegraphics[width=\columnwidth]{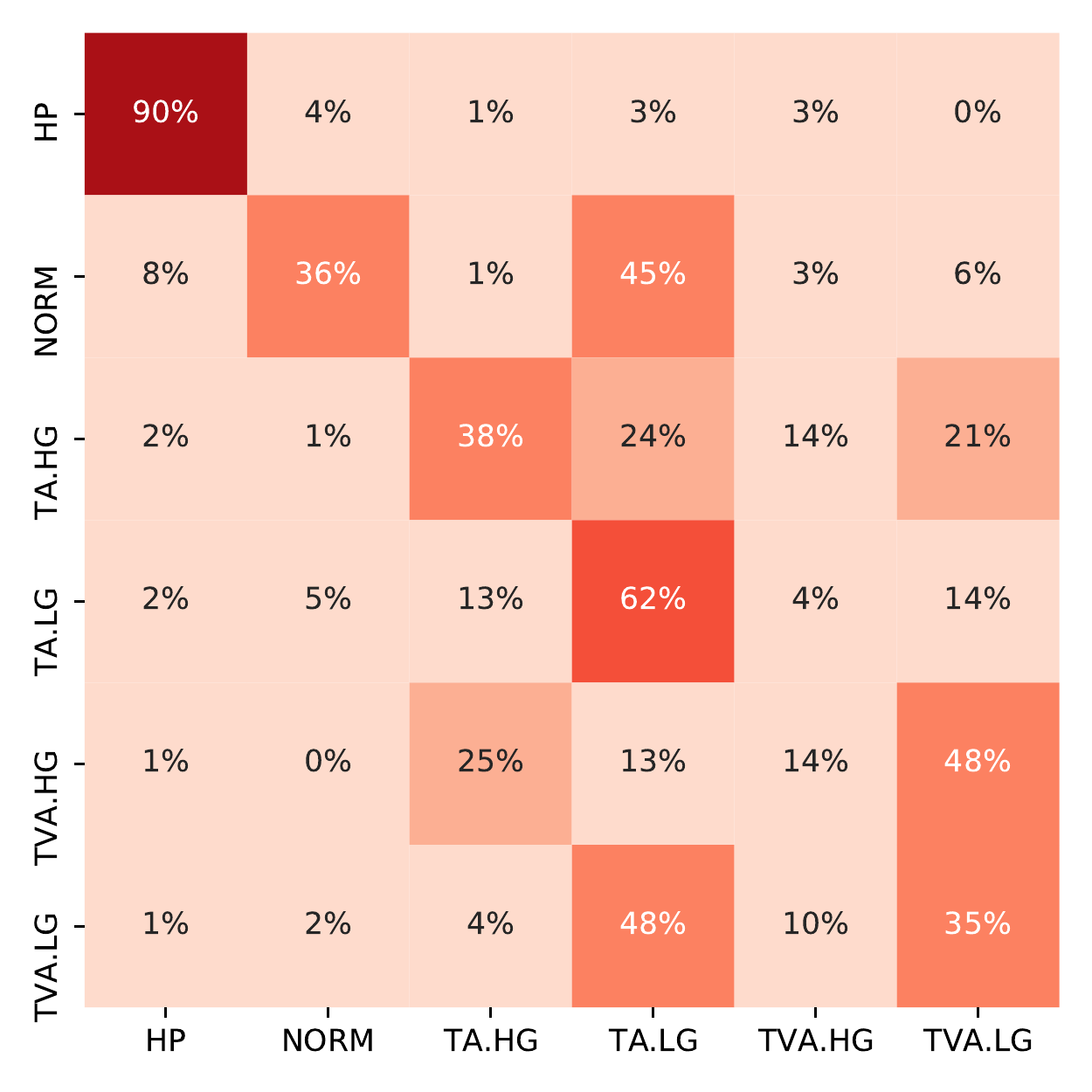}
            \caption{~ Baseline}
            \label{fig:cm-baseline}
        \end{subfigure}
    \end{minipage}
    \begin{minipage}[b]{.49\textwidth}
        \centering
        \begin{subfigure}{0.9\columnwidth}
            \centering
            \includegraphics[width=\columnwidth]{imgs/cm/multires-6c.pdf}
            \caption{~ Multi-resolution Ensemble}
            \label{fig:cm-ensemble}
        \end{subfigure}
    \end{minipage}
    \caption{Confusion matrices for the (a) baseline at ${\sigma=1500}$ (b) multi-resolution ensemble, reaching a BA of 0.46 and 0.67 respectively.}
    \label{fig:multires-6c}
\end{figure*}

%% file: 6_experiments.tex
\section{Experimental Results}

We evaluate our proposed method on the 7000$\times$7000\textmu m test patches of UniToPatho.
As a preliminary step, we tune the classifier responsible for the dysplasia grade prediction to find a suitable $T_d$ threshold.
In the following experiments we choose ${T_d = 0.2}$, as it strikes a favorable balance between false positives and sensitivity.
Lower thresholds may be preferred, for example, in clinical applications where minimizing the false negatives rate is more important.

Fig.~\ref{fig:multires-6c} shows the 6-class confusion matrix of the proposed method compared to the baseline approach described in Sec. \ref{method:preliminary}; in this latter case we employ ${\sigma=1500}$ which, as already discussed, yields the best overall BA.
Looking at the diagonal, it is quite evident that the proposed method significantly improves in average accuracy, that leaps from 0.46 to 0.67 (50\% relative increase). 
We notice how the baseline model is biased towards the lower grade classes: this represents further proof that subsampling the images results in the lack of useful features to distinguish high grade from lower grade and normal tissue. On the other hand, the multi-resolution approach shows remarkable improvements in assessing the correct grade. 
%
%
%

\addtolength{\tabcolsep}{-1pt}  
\begin{table}
    \centering
    \vspace{5pt}
    \begin{tabular}{l c c c c c c}
        \toprule
         & \ \  HP\ \ \   & NORM & \multicolumn{2}{c}{TA} & \multicolumn{2}{c}{TVA}\\
         & & & HG & LG & HG & LG \\
        \midrule
        Sensitivity	&0.86	&0.79	&0.60   &0.50	&0.78	&0.52 \\
        Specificity	&0.93   &0.87	&0.92   &0.94	&0.96	&0.92 \\
        BA	&0.89	&0.83	&0.76   &0.72	&0.87	&0.72 \\
        \bottomrule
        \end{tabular}
    \caption{ Sensitivity, Specificity and BA per class.}
    \label{tab:stats}
\end{table}
\addtolength{\tabcolsep}{1pt}  

\begin{table}
    \centering
    \begin{tabular}{r c c c c c}
        \toprule
        & $\sigma$ & HP & NORM & TA & TVA \\
        \midrule
        Baseline & 800 & \textbf{0.92} & 0.66 & 0.66 & 0.67 \\
        Baseline & 1500 & 0.85 & 0.72 & 0.65 & 0.68 \\
        Baseline & 7000 & 0.60 & 0.78 & 0.76 & 0.84 \\
        Multi-resolution & -  & 0.89 & \textbf{0.83} & \textbf{0.81} & \textbf{0.87} \\
        \bottomrule
    \end{tabular}
    \caption{Comparison of the class BA between the baseline and the proposed multi-resolution approach.}
    \label{tab:class-ba-compare}
\end{table}

Tab.~\ref{tab:stats} reports other metrics that are common in the related literature.
Our proposed method achieves quite high specificity for all classes, and we also observe promising sensitivity values especially for the higher-risk TA.HG and TVA.HG classes. 
%
%

Finally, we also analyze the per-type performance, as done in Sec.~\ref{method:preliminary}. The results are shown Tab.~\ref{tab:class-ba-compare}. Despite the small HP class accuracy drop - which could be due to the simple inference method - we observe an increase for all of the other tissue classes.
Notably, we obtain a great reduction in false positive adenoma predictions, and, most importantly, a more precise distinction between TA and TVA adenomas.



%% file: 7_conclusion.tex
\section{Conclusions}
\label{sec:conclusion}
%

This work presents UniToPatho, a histopathological dataset of colorectal polypss obtained from 292 high-resolution annotated images. UniToPatho provides annotation for hyperplastic polyps, adenomas (tubular or tubulo-villous) and their dysplasia grade (low or high).
We show that a single deep neural network fails at correctly classifying the tissue type.
We highlight that each of the classes discriminant features are extracted at different resolutions, making a direct classification of the tissue at a fixed scale a sub-optimal approach.
This observation allowed us to design a multi-resolution deep learning strategy that, employing an ensemble of classifiers, achieves 67\% accuracy. 
From a clinical perspective, the most relevant results are the differentiation capability between tubular and tubulo-villous adenomas and the dysplasia grading, which are the most difficult tasks for pathologists~\cite{gupta2020recommendations,hassan2020post,matsuda2016surveillance,rutter2020british}. 
The challenge to improve the automatic diagnostic accuracy is however still open, as UniToPatho is publicly available and still growing.
Our future work will focus on collecting samples from other institutions to assess the cross-laboratory generalization capability. 

%% file: 0_main.bbl
\begin{thebibliography}{10}

\bibitem{gonzalez2020updates}
Raul~S Gonzalez,
\newblock ``Updates and challenges in gastrointestinal pathology,''
\newblock {\em Surgical Pathology Clinics}, vol. 13, no. 3, pp. ix, 2020.

\bibitem{bevan2018colorectal}
Roisin Bevan and Matthew~D Rutter,
\newblock ``Colorectal cancer screening—who, how, and when?,''
\newblock {\em Clinical endoscopy}, vol. 51, no. 1, pp. 37, 2018.

\bibitem{denis2009diagnostic}
Bernard Denis, Carol Peters, Catherine Chapelain, Isabelle Kleinclaus, Anne
  Fricker, Richard Wild, Bernard Auge, Isabelle Gendre, Philippe Perrin, Denis
  Chatelain, et~al.,
\newblock ``Diagnostic accuracy of community pathologists in the interpretation
  of colorectal polyps,''
\newblock {\em European journal of gastroenterology \& hepatology}, vol. 21,
  no. 10, pp. 1153--1160, 2009.

\bibitem{mollasharifi2020interobserver}
Tahmineh Mollasharifi, Mahsa Ahadi, Elena Jamali, Afshin Moradi, Parisa
  Asghari, Saman Maroufizadeh, and Behrang Kazeminezhad,
\newblock ``Interobserver agreement in assessing dysplasia in colorectal
  adenomatous polyps: A multicentric iranian study,''
\newblock {\em Iranian Journal of Pathology}, pp. 167--174, 2020.

\bibitem{Janowczyk16}
Madabhushi~A Janowczyk~A,
\newblock ``Deep learning for digital pathology image analysis: A comprehensive
  tutorial with selected use cases,''
\newblock {\em Journal of pathology informatics}, pp. 7--29, 2016.

\bibitem{korbar2017deep}
Bruno Korbar, Andrea~M Olofson, Allen~P Miraflor, Catherine~M Nicka, Matthew~A
  Suriawinata, Lorenzo Torresani, Arief~A Suriawinata, and Saeed Hassanpour,
\newblock ``Deep learning for classification of colorectal polyps on
  whole-slide images,''
\newblock {\em Journal of pathology informatics}, vol. 8, 2017.

\bibitem{he2016deep}
Kaiming He, Xiangyu Zhang, Shaoqing Ren, and Jian Sun,
\newblock ``Deep residual learning for image recognition,''
\newblock in {\em Proceedings of the IEEE conference on computer vision and
  pattern recognition}, 2016, pp. 770--778.

\bibitem{wei2020evaluation}
Jason~W Wei, Arief~A Suriawinata, Louis~J Vaickus, Bing Ren, Xiaoying Liu,
  Mikhail Lisovsky, Naofumi Tomita, Behnaz Abdollahi, Adam~S Kim, Dale~C
  Snover, et~al.,
\newblock ``Evaluation of a deep neural network for automated classification of
  colorectal polyps on histopathologic slides,''
\newblock {\em JAMA Network Open}, vol. 3, no. 4, pp. e203398--e203398, 2020.

\bibitem{song2020automatic}
Zhigang Song, Chunkai Yu, Shuangmei Zou, Wenmiao Wang, Yong Huang, Xiaohui
  Ding, Jinhong Liu, Liwei Shao, Jing Yuan, Xiangnan Gou, et~al.,
\newblock ``Automatic deep learning-based colorectal adenoma detection system
  and its similarities with pathologists,''
\newblock {\em BMJ open}, vol. 10, no. 9, pp. e036423, 2020.

\bibitem{DeepHealth}
DeepHealth,
\newblock ``Deep-learning and hpc to boost biomedical applications for
  health,''
\newblock 2019.

\bibitem{unitopatho}
Luca Bertero; Carlo Alberto Barbano; Daniele Perlo; Enzo Tartaglione; Paola
  Cassoni; Marco Grangetto; Attilio Fiandrotti; Alessandro Gambella;~Luca
  Cavallo,
\newblock ``Unitopatho,'' \url{https://dx.doi.org/10.21227/9fsv-tm25}, 2021.

\bibitem{Tseung2005RobbinsAC}
J.~Tseung,
\newblock ``Robbins and cotran pathologic basis of disease: 7th edition,''
\newblock {\em Pathology}, vol. 37, pp. 190, 2005.

\bibitem{NCI}
National~Cancer Institute,
\newblock ``Nci dictionary of cancer terms - dysplasia,''
\newblock in {\em
  {https://www.cancer.gov/publications/dictionaries/cancer-terms/def/dysplasia}}.

\bibitem{kather2018}
Jakob~Nikolas Kather, Niels Halama, and Alexander Marx,
\newblock ``{100,000 histological images of human colorectal cancer and healthy
  tissue},''
\newblock Apr. 2018.

\bibitem{taherian2020tubular}
Mehran Taherian, Saran Lotfollahzadeh, Parnaz Daneshpajouhnejad, and Komal
  Arora,
\newblock {\em Tubular Adenoma},
\newblock StatPearls Publishing, Treasure Island (FL), 2020.

\bibitem{mehtat2020moleculary}
Mehtat Unlu, Evren Uzun, Goksel Bengi, Ozgul Sagol, and Sulen Sarioglu,
\newblock ``Molecular characteristics of colorectal hyperplastic polyp
  subgroups,''
\newblock {\em The Turkish Journal of Gastroenterology}, vol. 31, 05 2020.

\bibitem{he2015spatial}
Kaiming He, Xiangyu Zhang, Shaoqing Ren, and Jian Sun,
\newblock ``Spatial pyramid pooling in deep convolutional networks for visual
  recognition,''
\newblock {\em IEEE transactions on pattern analysis and machine intelligence},
  vol. 37, no. 9, pp. 1904--1916, 2015.

\bibitem{van2019evolutionary}
Gerard~Jacques van Wyk and Anna~Sergeevna Bosman,
\newblock ``Evolutionary neural architecture search for image restoration,''
\newblock in {\em 2019 International Joint Conference on Neural Networks
  (IJCNN)}. IEEE, 2019, pp. 1--8.

\bibitem{gupta2020recommendations}
Samir Gupta, David Lieberman, Joseph~C Anderson, Carol~A Burke, Jason~A
  Dominitz, Tonya Kaltenbach, Douglas~J Robertson, Aasma Shaukat, Sapna Syngal,
  and Douglas~K Rex,
\newblock ``Recommendations for follow-up after colonoscopy and polypectomy: a
  consensus update by the us multi-society task force on colorectal cancer,''
\newblock {\em Gastrointestinal endoscopy}, vol. 91, no. 3, pp. 463--485, 2020.

\bibitem{hassan2020post}
Cesare Hassan, Giulio Antonelli, Jean-Marc Dumonceau, Jaroslaw Regula, Michael
  Bretthauer, Stanislas Chaussade, Evelien Dekker, Monika Ferlitsch, Antonio
  Gimeno-Garcia, Rodrigo Jover, et~al.,
\newblock ``Post-polypectomy colonoscopy surveillance: European society of
  gastrointestinal endoscopy (esge) guideline--update 2020,''
\newblock {\em Endoscopy}, vol. 52, no. 08, pp. 687--700, 2020.

\bibitem{matsuda2016surveillance}
Takahisa Matsuda, Han-Mo Chiu, Yasushi Sano, Takahiro Fujii, Akiko Ono, and
  Yutaka Saito,
\newblock ``Surveillance colonoscopy after endoscopic treatment for colorectal
  neoplasia: from the standpoint of the asia--pacific region,''
\newblock {\em Digestive Endoscopy}, vol. 28, no. 3, pp. 342--347, 2016.

\bibitem{rutter2020british}
Matthew~D Rutter, James East, Colin~J Rees, Neil Cripps, James Docherty, Sunil
  Dolwani, Philip~V Kaye, Kevin~J Monahan, Marco~R Novelli, Andrew Plumb,
  et~al.,
\newblock ``British society of gastroenterology/association of coloproctology
  of great britain and ireland/public health england post-polypectomy and
  post-colorectal cancer resection surveillance guidelines,''
\newblock {\em Gut}, vol. 69, no. 2, pp. 201--223, 2020.

\end{thebibliography}
